\documentclass[epj]{webofc}
\usepackage[utf8]{inputenc}
\usepackage[varg]{txfonts}   
\usepackage{booktabs}
\usepackage{xcolor}
\definecolor{darkred}{rgb}{0.4,0.0,0.0}
\definecolor{darkgreen}{rgb}{0.0,0.4,0.0}
\definecolor{darkblue}{rgb}{0.0,0.0,0.4}
\usepackage[bookmarks,linktocpage,colorlinks,
    linkcolor = darkred,
    urlcolor  = darkblue,
    citecolor = darkgreen]{hyperref}
\usepackage{subfigure}
\usepackage{caption}
\wocname{EPJ Web of Conferences}
\woctitle{Lattice2017}
\begin{document}
%
\selectlanguage{english}
\title{%
Loop-TNR analysis of CP(1) model with theta term
}
\author{%
\firstname{Hikaru} \lastname{Kawauchi}\inst{1}\fnsep\thanks{Speaker, \email{kawauchi@hep.s.kanazawa-u.ac.jp} } \and
\firstname{Shinji} \lastname{Takeda}\inst{1}
}
\institute{%
Institute for Theoretical Physics, Kanazawa University, Kanazawa 920-1192, Japan
}
\abstract{%
The phase structure of the two dimensional lattice CP(1) model in the presence of the $\theta$ term is analyzed by tensor network methods. 
The tensor renormalization group, which is a standard renormalization method of tensor networks, is used for the regions $\theta =0$ and $\theta \neq 0$. Loop-TNR, which is more suitable for the analysis of near criticality, is also implemented for the region $\theta =0$. The application of Loop-TNR for the region $\theta \neq 0$ is left for future work.
}
\maketitle
\section{Introduction}\label{intro}

Haldane conjecture implies that the two dimensional O(3) nonlinear sigma model with $\theta = \pi$ is gapless \cite{Haldane1983_1, Haldane1983_2, Haldane1985, Haldane2016, Affleck1985}. There are several Monte Carlo studies for the O(3) model around $\theta =\pi$ \cite{Bietenholz1995, Alles2007, Bogli2012, Forcrand2012, Azcoiti2012, Alles2014}. Those results confirm the critical behavior. Furthermore, the critical exponent and the exponent of the logarithmic correction expected in Refs.~\cite{Affleck1987, Affleck1989} are verified too. In Ref.~\cite{Azcoiti2007}, the phase structure of the two dimensional CP(1) model is studied and they conclude that there is a second order phase transition line at $\theta = \pi$. However, the critical exponent of the CP(1) model is not agreement with that of the O(3) model and changes continuously in spite of the equality of those two models in the continuum limit.

Our purpose is to reanalyze the phase structure of the CP(1) model by using sign-problem-free methods, namely the tensor renormalization group (TRG) \cite{Levin2006} and the loop optimization for tensor network renormalization (Loop-TNR) \cite{Shuo2017}. 


\section{A brief review of Loop-TNR}\label{LoopTNR}

\begin{figure}[t] 
  \centering
  \includegraphics[width=0.8\textwidth,clip,bb=0 0 900 350]{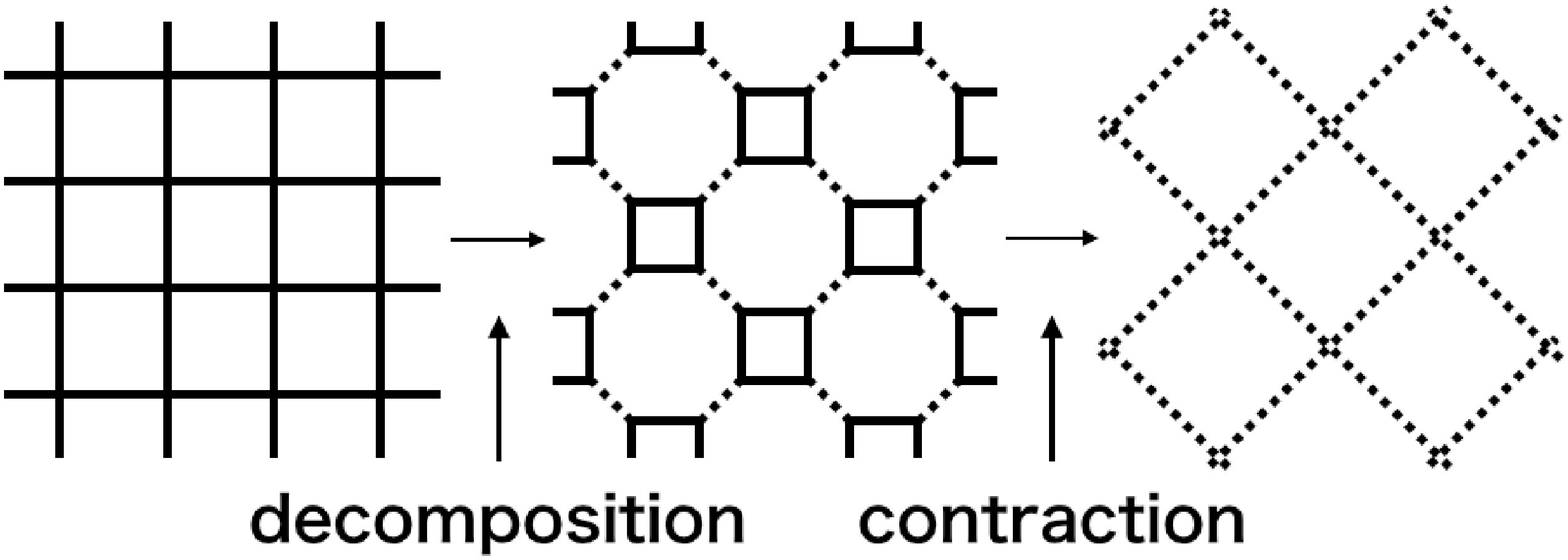}
  \caption{Renormalization of tensor network is divided into mainly two steps, (II) decomposition of tensors and (III) contraction of tensors. The calculation of (III) can be done exactly, but that of (II) includes some errors. }
   \label{TRG}
\end{figure}

In this section, we briefly explain the procedure of Loop-TNR\cite{Shuo2017}, whose main steps are almost the same as TRG\cite{Levin2006}. The both methods are divided into three steps, (I) construction of a tensor network representation of what one want to calculate, e.g. a partition function of the target system, and two coarse-graining steps, (II) decomposition of the tensors and (III) contraction of the indices of the tensors. Implementing the steps (II) and (III) iteratively, the number of tensors decreases and one can finish the computation approximately. Here, we focus on the latter two steps, which are illustrated in Fig.~\ref{TRG}. The difference between TRG and Loop-TNR is present in the step (II) while the step (III) is the same in the both methods. We explain the difference below.

\begin{figure}[thb]
\centering
  \includegraphics[width=0.6\textwidth,clip,bb=0 0 900 200]{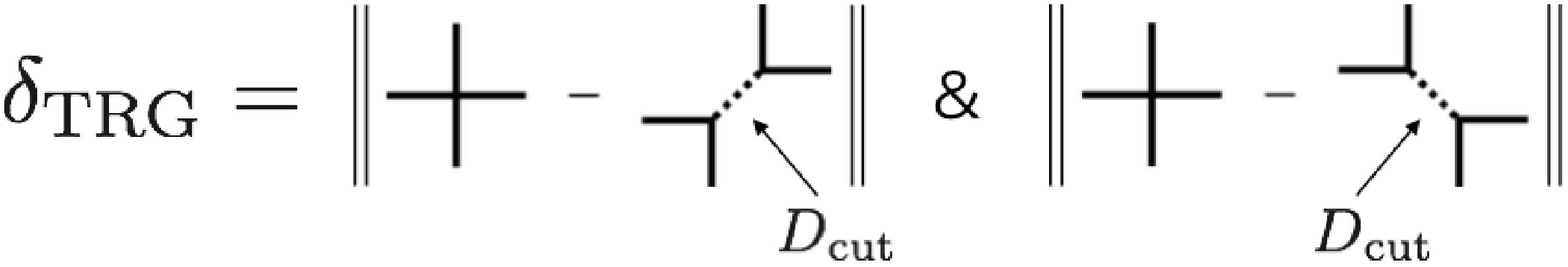}
  \caption{Cost function of TRG.}
  \label{cost_trg}
\end{figure}

\begin{figure}[t]
   \subfigure[Relative errors of the partition function as a function of the bond dimension $D_{\rm cut}$ at the critical point.]
             {\includegraphics[width=0.475\textwidth,clip,bb=0 0 405 300]{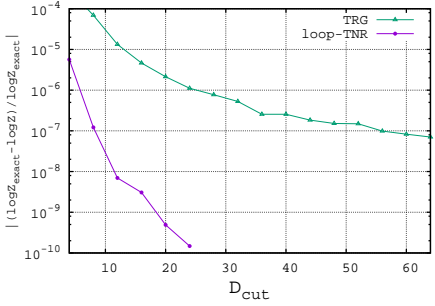}}\hfill
   \subfigure[Relative errors of the partition function as a function of the number of the loop optimization $N_{\rm opt}$ at the critical point.]%
             {\includegraphics[width=0.475\textwidth,clip,bb=0 0 405 300]{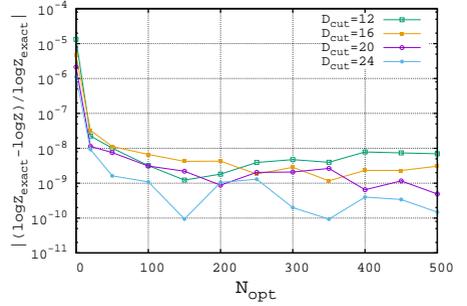}}
   \caption{An example of implementation of TRG and Loop-TNR. The figures show the relative errors of the partition function of the two dimensional Ising model at the critical temperature. The number of spins is $2^{39}$.}
   \label{logz}
\end{figure}

In TRG, the tensors in the networks are decomposed by using the singular value decomposition (SVD), and the discard of the some small singular values makes the numerical computation feasible. The decomposition using SVD corresponds to the minimization of the cost function $\delta_{\rm TRG}$ described in Fig.~\ref{cost_trg}, where the dotted lines have the degrees of freedom equal to the number of the remained singular values. We refer to the bond dimension as $D_{\rm cut}$. Thus, the larger $D_{\rm cut}$ value one takes, the smaller the error coming from TRG algorithm is. 
Here, we show an example of results of TRG in Fig.~\ref{logz}~(a). The triangle dots indicate the relative errors of the partition function of the two dimensional Ising model at the critical temperature versus $D_{\rm cut}$.  As indicated in the figure, the slope is not so steep. In fact, TRG is not suitable for the analysis of critical region and the error of TRG near criticality becomes large compared to that of off critical region \cite{Xie2012}. Therefore, even if one sets $D_{\rm cut}$ to large number, the error does not become small so much.

The reason why TRG does not work so well near critical point was discussed in Refs.~\cite{Gu2009, Evenbly2015}, where it is insisted that TRG does not renormalize short-range correlation properly. In order to overcome the deficit of TRG algorithm, Evenbly and Vidal newly developed the tensor network renormalization (TNR), which can properly renormalize short-range correlation even in critical region \cite{Evenbly2015}.

\begin{figure}[thb] 
\centering
  \includegraphics[width=0.5\textwidth,clip,bb=0 0 700 300]{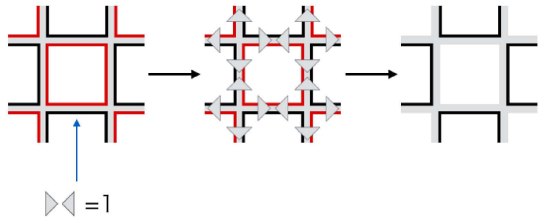}
  \caption{Entanglement filtering step.}
  \label{ef}
\end{figure}

\begin{figure}[thb] 
\centering
  \includegraphics[width=0.5\textwidth,clip,bb=0 0 900 200]{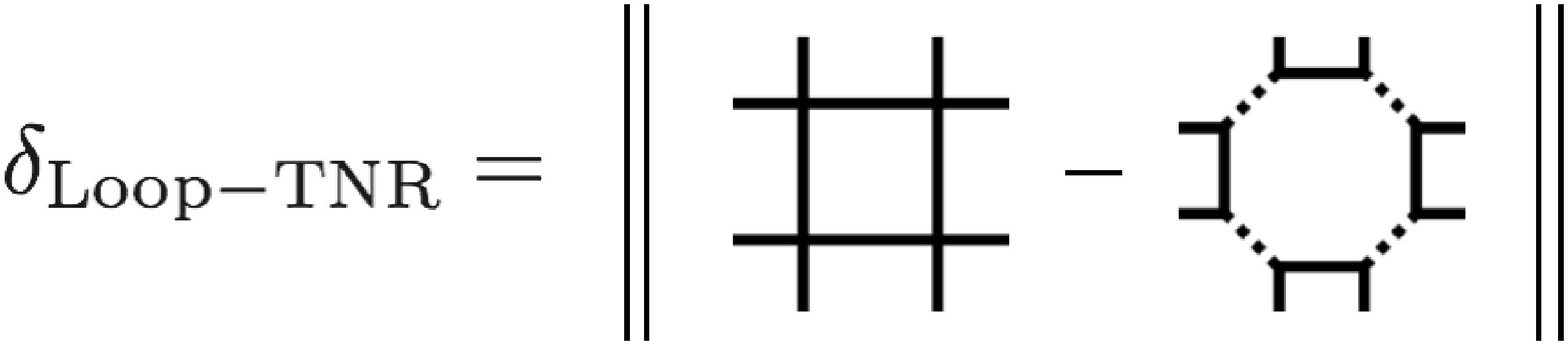}
  \caption{Cost function of Loop-TNR.}
  \label{cost_looptnr}
\end{figure}

Loop-TNR is one of the alternatives to TNR. In this method, (II) decomposition step in Fig.~\ref{TRG} is divided into two steps, (i) entanglement filtering and (ii) loop optimization. In (i) entanglement filtering step, some projectors are inserted between the tensors and deform the corner double line (CDL) tensors \cite{Gu2009, Evenbly2015} as shown in Fig.~\ref{ef}. The CDL tensors contain only short-range correlations and can not be renormalized properly by TRG method. If the system contains the CDL tensors, projectors can reduce the bond dimension. We skip the detail of this step. In (ii) loop optimization step, the cost function of TRG $\delta_{\rm TRG}$ is replaced by that of Loop-TNR $\delta_{\rm Loop-TNR}$ in Fig.~\ref{cost_looptnr}. The initial eight tensors in the octagonal tensor network are prepared by using SVD, and the eight tensors are updated in turn site-by-site. 
By repeating this procedure, one can obtain the optimized tensors.
The combination of (i) and (ii) can renormalize the short-range correlations properly. Figure~\ref{logz}(b) shows that the relative errors become smaller as $N_{\rm opt}$ becomes larger, where $N_{\rm opt}$ denotes the number of the optimization on a loop. When $N_{\rm opt}$ is large enough, the errors become small drastically as shown in Fig.~\ref{logz}(a) for fixed $D_{\rm cut}$.

\section{Numerical results}\label{results}

We apply these methods to the CP(1) model and show the numerical results below. TRG method is used for the both cases $\theta = 0$ and $\theta \neq 0$, and Loop-TNR method is used only for the case $\theta = 0$. We use the tensor network representation of the CP(1) model in Ref.~\cite{Kawauchi2016}. The tensor $T_{{\rm CP}(1)}$ can be described by the combination of the two tensor $T'(\beta)$ and $T''(\theta)$,
\begin{eqnarray}
T_{{\rm CP}(1)}=T'(\beta)\otimes T''(\theta).
\end{eqnarray}
We truncate initially the bond dimension of the tensor $T'(\beta)$ to some value $D_{\beta}$ and $T''(\theta)$ to $D_{\theta}$, that is, the total bond dimension of the initial tensor $T_{{\rm CP}(1)}$ is $D_{\beta}\times D_{\theta}$. 
This is reasonable since the absolute values of the elements of the tensors decrease monotonically as a function of the absolute value of the each index.
And we fix the bond dimensions of the renormalized tensors  to $D_{\beta}\times D_{\theta}$ at each renormalization step.

\subsection{Application of TRG and Loop-TNR to CP(1) model without the $\theta$ term}\label{LoopTNRtoCP1}

TRG and Loop-TNR are applied to the CP(1) model at $\theta = 0$. By using those methods, we calculate the partition function of the CP(1) model. And we define the specific heat as
\begin{eqnarray}
C=\frac{\beta^2}{L^2}\frac{\partial^2 {\rm log}Z}{\partial \beta^2}.
\end{eqnarray}

\begin{figure}[htb]
  \hspace{15mm}
  \includegraphics[width=0.7\textwidth,clip,bb=0 0 405 305]{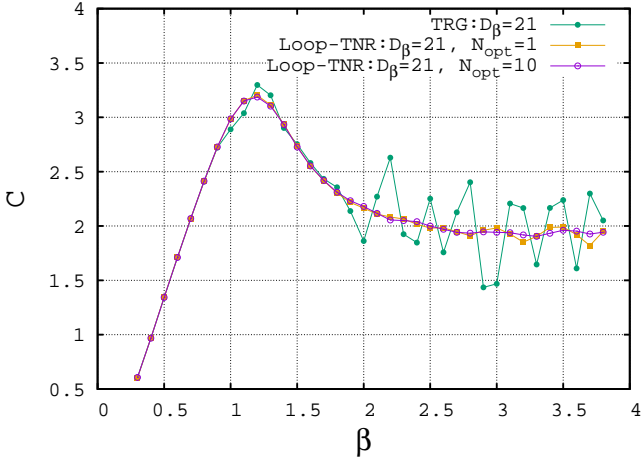}
  \vspace{-10mm}
  \caption{Specific heat of the CP(1) model at $\theta = 0$ as a function of $\beta$. The linear lattice size is $L=2^{20}$. The fluctuation coming from the error due to the TRG algorithm becomes eventually smaller as $N_{\rm opt}$ is larger.}
  \label{C_cp1_loopTNR}
\end{figure}

We take the derivative numerically and Fig.~\ref{C_cp1_loopTNR} shows the result of $C$. The bond dimension is fixed at $D_{\beta}=21$ and the linear lattice size is $L=2^{20}$. The number of the loop optimization in Loop-TNR is $N_{\rm opt}=1$ and $10$. If the error of the partition function is large, the result of the numerical derivation with respect to $\beta$ fluctuates. As can be seen from this figure, the fluctuation becomes gradually smaller as the value $N_{\rm opt}$ is larger. This result suggests that the loop optimization makes the error due to the TRG algorithm small.

\subsection{Application of TRG to CP(1) model with the $\theta$ term}\label{TRGtoCP1}

\begin{figure}[htb] 
  \hspace{20mm}
  \includegraphics[width=0.6\textwidth,clip,bb=0 0 405 305]{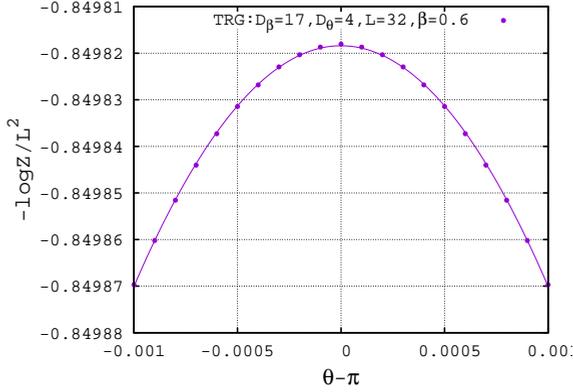}
  \vspace{-10mm}
  \caption{$-\frac{1}{L^2}{\rm log}Z$ of the CP(1) model at $\beta = 0.6$ as a function of $\theta$. The linear lattice size is $L=32$. The dots are the results of TRG fitted by the curved line of a polynomial fitting.}
  \label{logz_cp1}
\end{figure}

Next, we show the results of CP(1) model with the $\theta$ term obtained by TRG.
By using this method, the partition function $Z$ can be computed approximately. Figure~\ref{logz_cp1} is the result of $-\frac{1}{L^2}{\rm log}Z$ at $\beta = 0.6$ as a function of $\theta$, where the linear lattice size is $L=32$. The dots are calculated by the TRG method of $D_{\beta} =17$ and $D_{\theta} =4$. The curved line is drawn by a polynomial interpolation,
\begin{eqnarray}\label{logz_fit}
-\frac{1}{L^2}{\rm log}Z = c_0 + c_2 (\theta - \pi)^2 + c_4 (\theta - \pi)^4 + \cdots ,
\end{eqnarray}
where $c_0$, $c_2$ and $c_4$ are fitting parameters.

\noindent
We define the topological susceptibility as
\begin{eqnarray}\label{kai}
\chi(\theta)=\frac{1}{L^2}\frac{\partial^2 {\rm log}Z}{\partial \theta^2}.
\end{eqnarray}
From Eq.~\ref{logz_fit} and Eq.~\ref{kai}, the maximal value of $\chi(\theta)$ is easily obtained,
\begin{eqnarray}
\chi_{\rm max} = \chi(\theta = \pi)=-2c_2.
\end{eqnarray}

\begin{figure}[htb] 
  \hspace{25mm}
  \includegraphics[width=0.6\textwidth,clip,bb=0 0 405 305]{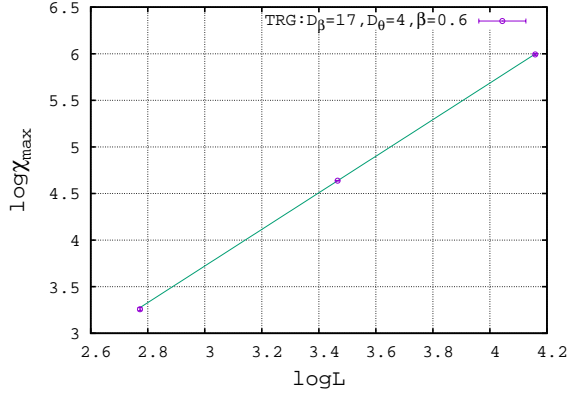}
  \vspace{-10mm}
  \caption{The volume dependence of $\chi_{\rm max}$ at $\beta = 0.6$.}
  \label{kaimax}
\end{figure}

\noindent
The order of the phase transition can be verified by the volume dependence of $\chi_{\rm max}$,
\begin{eqnarray}
\chi_{\rm max}\propto L^{b}.
\end{eqnarray}
The exponent $b$ can be obtained from the slope in Fig.~\ref{kaimax}

\begin{figure}[htb]
  \hspace{-5mm}
  \centering
  \includegraphics[width=0.6\textwidth,clip,bb=0 0 405 305]{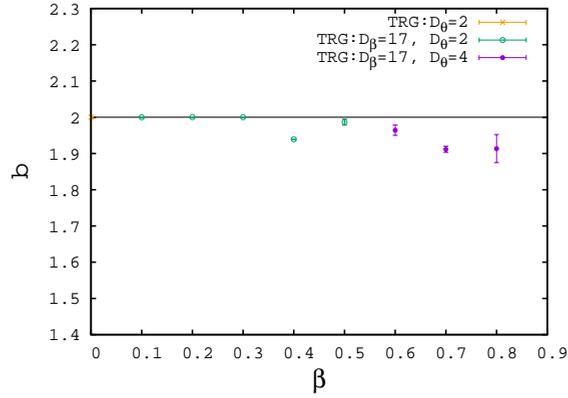}
  \vspace{-10mm}
  \caption{The exponent $b$ as a function of $\beta$. The bond dimensions $D_{\beta}$ and $D_{\theta}$ are chosen as the minimum values that make the fit in Fig.~\ref{logz_cp1} and \ref{kaimax} possible.}
  \label{gamma_over_nu}
\end{figure}

We carry out this procedure at various $\beta$s. The results are shown in Fig.~\ref{gamma_over_nu}. In the region $0\leq \beta \leq 0.3$, $b$ is almost $2$, which means the first order phase transition, while in the region $0.4\leq \beta \leq 0.8$, $b$ is less than $2$, which implies the second order phase transition. This tendency is almost consistent with the previous study\cite{Azcoiti2007}. Note that this result does not include the systematic errors due to the bond truncation $D_{\rm cut}$. 


%

\section{Summary}\label{summary}
In this report, we apply TRG and Loop-TNR to the two dimensional lattice CP(1) model without the $\theta$ term and confirm the effectiveness of Loop-TNR.
And the phase structure of the CP(1) model with the $\theta$ term is analyzed by using TRG. The tendency is confirmed that the order of the phase transition at $\theta =\pi$ is the first order for $\beta \leq 0.3$ and the second order for $0.4 \leq \beta$. For more precise study, we shall apply Loop-TNR to the region $\theta \neq 0$ and raise the bond dimension for future work.

\section*{Acknowledgments}
This work was supported by JSPS KAKENHI Grant Numbers JP17J03948, JP17K05411.

\clearpage
\bibliography{lattice2017}

\end{document}